\documentclass[10pt,journal,onecolumn,draftclsnofoot]{IEEEtran}

\usepackage{amsthm}
\usepackage{amsmath}
\emergencystretch=\maxdimen
\usepackage{amssymb}

\usepackage{orcidlink}
\usepackage{subfig}

\usepackage{csquotes}
\newtheorem{theorem}{Theorem}[section]
\newtheorem{definition}[theorem]{Definition}
\newtheorem{remark}[theorem]{Remark}
\newtheorem{example}[theorem]{Example}

\DeclareMathOperator{\R}{\mathbb{R}}
\newcommand{\Rn}{\mathbb{R}_{> 0}}

\DeclareMathOperator{\len}{len}
\DeclareMathOperator{\dist}{dist}
\DeclareMathOperator{\occ}{occ}
\DeclareMathOperator{\iv}{iv}
\DeclareMathOperator{\tr}{tr}
\DeclareMathOperator{\rg}{rg}
\DeclareMathOperator{\bd}{bd}
\DeclareMathOperator{\tim}{tim}
\newcommand{\tl}{\underline{t}}
\newcommand{\tu}{\overline{t}}
\newcommand{\al}{\underline{\alpha}}
\newcommand{\au}{\overline{\alpha}}
\newcommand{\dl}{\underline{\delta}}
\newcommand{\du}{\overline{\delta}}

\usepackage{tabularx, environ}
\usepackage{changepage}
\makeatletter
\newcommand{\problemtitle}[1]{\gdef\@problemtitle{#1}}%
\newcommand{\probleminput}[1]{\gdef\@probleminput{#1}}%
\newcommand{\problemquestion}[1]{\gdef\@problemquestion{#1}}%
\newlength{\problemindent} %
\NewEnviron{problem}{%
	\problemtitle{}\probleminput{}\problemquestion{}%
	\BODY%
	\par\addvspace{.5\baselineskip}
	\noindent
	\hspace{\problemindent}\centering{\textsc{\@problemtitle}} %
	\begin{adjustwidth}{\problemindent}{\problemindent} %
		\setlength{\itemindent}{\dimexpr\labelwidth+\problemindent} %
		\begin{itemize}
			\item[\textbf{Input:}] \@probleminput
			\item[\textbf{Task:}] \@problemquestion
		\end{itemize}
	\end{adjustwidth}
	\par\addvspace{.5\baselineskip}
}
\makeatother

\graphicspath{{fig}}

\usepackage[
type={CC},
modifier={by-nc-nd},
version={4.0},
]{doclicense}

\newcommand\copyrighttextmanual{%
	\footnotesize \textcopyright 2025. This manuscript version is made available under the CC-BY-NC-ND 4.0 license \url{https://creativecommons.org/licenses/by-nc-nd/4.0/}
}
\newcommand\copyrightnoticemanual{%
	\begin{tikzpicture}[remember picture,overlay]
		\node[anchor=south, yshift=10pt] at (current page.south) {
				\begin{minipage}[c]{\dimexpr\textwidth+8em\relax}
					\begin{minipage}[c]{0.86\linewidth}
						\copyrighttextmanual
					\end{minipage}%
					\hfill
					\begin{minipage}[c]{0.14\linewidth}
						\centering
						\doclicenseImage[imagewidth=7em]
					\end{minipage}
				\end{minipage}
		};
	\end{tikzpicture}%
}

\begin{document}
	
	\tikzstyle{vssborder} = [dot diameter=0.9pt, dot spacing=1.5pt, dots]	
	\usetikzlibrary{patterns}
	
	\title{Design Tasks and Their Complexity\\for the European Train Control System\\with Hybrid Train Detection}
	\author{Stefan~Engels\,\orcidlink{0000-0002-0844-586X},
		Tom~Peham\,\orcidlink{0000-0003-3434-7881},
		Judith~Przigoda\,\orcidlink{0009-0003-3724-9205},
		Nils~Przigoda\,\orcidlink{0000-0001-8947-3282},
		and~Robert~Wille\,\orcidlink{0000-0002-4993-7860},%
		\IEEEcompsocitemizethanks{\IEEEcompsocthanksitem S. Engels and T. Peham are with the Technical University of Munich, TUM School of Computation, Information and Technology, Chair for Design Automation, 80333 Munich, Germany.\protect\\
			E-mail: \{stefan.engels, tom.peham\}@tum.de
			\IEEEcompsocthanksitem J. Przigoda and N. Przigoda are with Siemens Mobility GmbH, 38126 Braunschweig, Germany.\protect\\
			E-mail: \{judith.przigoda,nils.przigoda\}@siemens.com
			\IEEEcompsocthanksitem R. Wille is with the Technical University of Munich, TUM School of Computation, Information and Technology, Chair for Design Automation, 80333 Munich, Germany, and also with the Software Competence Center Hagenberg GmbH (SCCH), 4232 Hagenberg, Austria.\protect\\
			E-mail: robert.wille@tum.de}%
	}

	\IEEEtitleabstractindextext{%
		\begin{abstract}
	Railway networks have become increasingly important in recent times, especially in moving freight and public transportation from road traffic and planes to more environmentally friendly trains. Since expanding the global railway network is time- and resource-consuming, maximizing the rail capacity of the existing infrastructure is desirable. However, simply running more trains is infeasible as certain constraints enforced by the train control system must be satisfied. The capacity of a network depends (amongst others) on the distance between trains allowed by this safety system. While most signaling systems rely on fixed blocks defined by costly hardware, new specifications provided by Level~2 with Hybrid~Train~Detection of the European Train Control System~(ETCS~L2~HTD), formerly known as ETCS~Hybrid~Level~3, allow the usage of virtual subsections. This additional degree of freedom allows for shorter train following times and, thus, more trains on existing railway tracks.
	
	On the other hand, new design tasks arise on which automated methods might be helpful for designers of modern railway networks. However, although first approaches exist that solve design problems arising within ETCS~L2~HTD, neither formal descriptions nor results on the computational complexity of the corresponding design tasks exist. In this paper, we fill this gap by providing a formal description of design tasks for ETCS~L2~HTD and proof that these tasks are NP-complete or NP-hard, respectively. By that, we are providing a solid basis for the future development of methods to solve those tasks, which will be integrated into the \emph{Munich Train Control Toolkit} available open-source on GitHub at \url{https://github.com/cda-tum/mtct}.
\end{abstract}
		
		}

	\maketitle
	\copyrightnoticemanual

	\IEEEdisplaynontitleabstractindextext

	\IEEEpeerreviewmaketitle

	\section{Introduction}\label{sec:intro}
Railway networks are an important part of both public transportation and logistics, e.g., to reduce carbon emissions compared to cars and planes. Unlike road traffic, trains have relatively long braking distances. Hence, trains cannot operate on sight due to safety constraints. Moreover, switches have to be safely set for trains to turn. Because of that, a control system is needed for efficient and safe rail traffic. Signaling systems are essential to allow or deny a train to use a specific track part. For this, many national train control systems have been developed. The European Union Agency for Railways has listed about 40 systems alone in Europe \cite{Gemine2023}. In Germany, for example, \emph{PZB~90} (intermittent train protection) and \emph{LZB} (continuous train control) are widely used \cite{Maschek2018}. For reasons of interoperability, these national control systems are unified in the \emph{European Train Control System} (ETCS) \cite{EuropeanCommission2024,Schnieder2021}. It is specified in the \emph{Control Command and Signalling Technical Support Instrument}~(CCS~TSI) \cite{EuropeanCommission2023}, in particular the set of specifications listed in its appendix A \cite{EUAR2023}. Even for metro lines, a radio-based system analog to ETCS (Level 2) is being implemented, namely the \emph{Communication Based Train Control} (CBTC) \cite{Schnieder2021a, SMG2019}. Similarly, major standardized systems have been implemented in China (\emph{Chinese Train Control System}, CTCS) and North America (\emph{Positive Train Control}, PTC) \cite{Pachl2021}.

\subsection{European Train Control System}
Currently, most railway systems in the world rely on block signaling, where the whole railway network is divided into blocks equipped with means of train detection to determine whether a given block is currently occupied by a train. Within a block, at most, there can only be one train at a time \cite{Pachl2019}. This paper focuses on ETCS, even though the model and key components likely carry over to other modern control systems. ETCS can be implemented in various levels  \cite{EuropeanCommission2024, Schnieder2021}.
Previously, the specifications distinguished three levels, however with the CCS TSI 2023 \cite{EuropeanCommission2023} Level~2 and Level~3 have been combined.

\begin{figure}[!t]
	\centering{
		\subfloat[ETCS Level 1]{\includegraphics[width=0.8\textwidth]{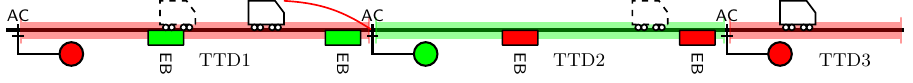}\label{fig:etcs-level-1}}
		\vspace{-5mm}
		\newline
		\subfloat[ETCS Level 2]{\includegraphics[width=0.8\textwidth]{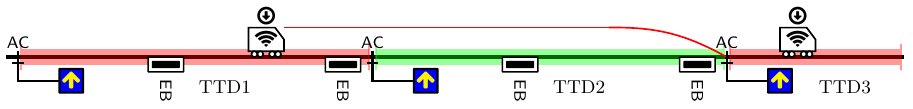}\label{fig:etcs-level-2}}
		\vspace{-5mm}
		\newline
		\subfloat[ETCS Level 2 Moving Block (formerly known as ETCS Level 3)]{\includegraphics[width=0.8\textwidth]{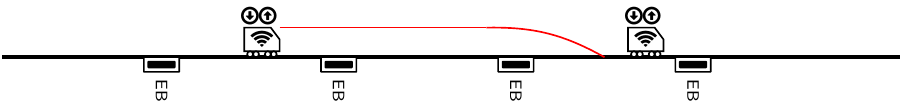}\label{fig:etcs-level-3}}
		\vspace{-5mm}
		\newline
		\subfloat[ETCS Level 2 Hybrid Train Detection (formerly known as ETCS Hybrid Level 3)]{\includegraphics[width=0.8\textwidth]{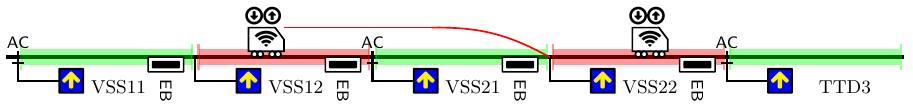}\label{fig:etcs-hybrid-level-3}}
		\vspace*{1mm}
		\caption{Schematic drawings of various ETCS levels by Engels et al. \cite{Engels2023} published under a CC BY 4.0 license}
		\label{fig:etcs-levels}
		\vspace*{1mm}
	}
\end{figure}

\emph{Level 1} (L1) relies on signals and hardware on the tracks (e.g., Eurobalises and Euroloops), which transmit position and signaling information at discrete points (Eurobalise) or predefined intervals (Euroloop). The trackside fully handles the detection of block occupations (e.g., using axle counters), see Figure~\ref{fig:etcs-level-1}, in which the following train slows down because it receives the moving authority into TTD2 only at the next Eurobalise.

In \emph{Level 2} (L2), a permanent GSM-R connection between the train and the control system exists. By doing so, the train can continuously receive the maximal permitted speed. Signals are now optional but are often included as a fallback system. The train receives its moving authority through the GSM-R connection and is no longer bound to the discrete positions of Eurobalises.

At the same time, train location and integrity can be managed at different parts of the control system.
Classically, reporting block sections as free works as in L1. However, due to the difference mentioned above in transmitting the moving authority, the trains in Figure~\ref{fig:etcs-level-2} can follow each other more closely.

The block section borders are equipped with \emph{Trackside Train Detection} (TTD) systems, e.g., axle counters, called TTD~sections. Thus far, those blocks have often been defined by geography and economic considerations or as a trade-off between efforts to install axle counters and possible benefits. Because of this, the length of the corresponding TTD sections ranges from some meters (e.g., around turnouts) or some hundred meters (e.g., at train stations) to several kilometers (e.g., in remote areas). Of course, this significantly affects the efficiency of the underlying railway networks.

With the introduction of \emph{Moving Block},  formerly known as ETCS~Level~3, those principles have changed for the first time since the 19th century. The train reports the exact position itself, and no trackside detection is necessary. For this, trains must be equipped with an integrity system that detects if a train is still complete. Only in this case can it safely release parts of the track. 	Rather than relying on fixed blocks, trains can follow at their braking distances since no hardware predefines any blocks, see Figure~\ref{fig:etcs-level-3}. However, in practice, such systems impose new challenges \cite{Bartholomeus2018}.

Hence, Level 2 can also be used with a compromise of both classical block signaling and moving blocks. In \emph{Level~2~Hybrid Train Detection}~(L2 HTD), formerly known as Hybrid~Level~3, TTD sections that already exist are divided into smaller \emph{Virtual Subsections} (VSS), which do not require physical axle counters anymore and, hence, allow for a much higher degree of freedom in the utilization of existing railway networks, see Figure~\ref{fig:etcs-level-3}. Independently of trackside hardware, the occupation of these virtual subsections is observed by the position and integrity data obtained from the trains \cite{EEUG2022}.

\subsection{Related Work}
This additional degree of freedom allows for shorter train following times and, thus, more trains on existing railway tracks. At the same time, new design tasks arise that require automated methods to be helpful for designers of modern railway networks.
However, thus far, design automation methods for ETCS, on the one hand, mainly focus on blocks characterized by hardware (e.g., axle counters). Because of this, they use general measures of capacity, energy efficiency, or economic efficiency independently from a specific schedule to propose a signaling layout \cite{Gill1992, Chang1998, Chang1999, Ke2005, Ke2009, Dillmann2019, Luteberget2019a, Vignali2020, Sun2020}. Thus, the respective solution is expected to work well on various schedules.
On the other hand, train routing was considered separately, assuming the block layout had already been decided upon. Approaches are mainly separated in macroscopic routing \cite{Goossens2006} and track allocation on a microscopic level, i.e., considering the constraints imposed by the control system and sections. Track allocation was considered in general \cite{DAriano2007, Borndoerfer2008, Lusby2009} and specifically focusing on railway stations  \cite{Carey2003, Lusby2013, Wang2022}.

This research has been extended to future signaling systems, such as ETCS L2 HTD and Moving Block control systems.
Headways on such railway lines are different to former theory leading to adaptions in the blocking-time theory \cite{Bueker2019}.
Besides that, recent research by Schlechte et al. optimizes time tables on dedicated moving block systems \cite{Schlechte2022}, which was later extended to use lazy constraints to speed up computation \cite{Klug2022, Engels2024FedCSISPreprint}.
However, implementing a moving block system in practice is significantly harder than using virtual subsections, in which case only the design but not the control logic changes \cite{Bartholomeus2018}.

While (almost) all of these methods can also be applied for ETCS~L2~HTD, because virtual subsections and TTD sections have essentially the same consequences in final operation (assuming no system fails), they are not utilizing the additional degree of freedom introduced by virtual subsections. Using ETCS~L2~HTD, changing a previous block layout defined by TTD~sections even \emph{after} building the network is now possible. In particular, it might be reasonable to consider a desired schedule directly because changing a block layout with a new schedule is more manageable than before. However, corresponding joint considerations of infrastructure planning and scheduling are just at the beginning.

For ETCS~L2~HTD, specifically, the contributions to the \enquote{Hybrid ERTMS/ETCS Level~3 Case Study} provided an step in this direction \cite{Butler2018}.
Here, the main goal was to detect potential ambiguities in the specification to fix them in later versions. Submitted papers do this using iUML-B, Event-B, Formal B, Electrum, SysML/KAOS, and SPIN \cite{Butler2018}. However, while these formalizations provided initial (non-ambiguous) interpretations of the ETCS~L2~HTD, no precise design tasks, e.g., how to generate new layouts, can be derived from that.
To the best of our knowledge, only research proposed in~\cite{Wille2021, Peham2022, Engels2023, Engels2024LBR} explicitly considered the automatic design of hybrid train detection systems based on specific timetable requests. However, even those papers specified the design tasks at a relatively high level, and neither formal descriptions nor results on the computational complexity of those tasks have been provided. 

Overall, all related work on developing design methods for ETCS~L2~HTD has been conducted without a clear definition nor a formal understanding of the complexity of the underlying task.
As a consequence, we currently have heuristics that usually generate non-optimal\footnote{with respect to some objective describing the operational outcome, see also Section~\ref{sec:tasks}} results, on the one hand, \cite{Peham2022} and exact solutions which significantly suffer from huge run-times and severely limited scalability on the other \cite{Wille2021, Engels2023, Engels2024LBR}, but do not know whether this is caused by insufficient methodology or complexity reasons. Similar scheduling problems (e.g., \cite{DAriano2007}) have been proven to be NP-hard by using the alternative graph introduced in \cite{Mascis2002}. At the same time, we are not aware of formal complexity results on the design tasks considered in~\cite{Wille2021, Peham2022, Engels2023, Engels2024LBR}.

In this paper, we aim to shed light on this. To this end, we first introduce a graph model and review the relevant concepts of ETCS~L2~HTD on this in Section~\ref{sec:preliminaries}. Afterward, we discuss three essential design tasks resulting from the new degrees of freedom and provide clear, i.e., formal, problem definitions and examples in Section~\ref{sec:tasks}. This includes proofs showing that these new design tasks remain complex, i.e., are NP-complete or NP-hard, respectively (even if some restrictions are added).
By this, we provide a proper and valuable theoretical foundation (both on the problem description and its complexity) for upcoming solution-oriented research in this area. 
Arising methods for the automated design of such systems will be made available within the \emph{Munich Train Control Toolkit} (MTCT) hosted on GitHub (\url{https://github.com/cda-tum/mtct}).

\section{Preliminaries}\label{sec:preliminaries}

In this section, we provide a basic definition of railway networks using graph terminology, which, step by step, is extended to cover the relevant properties. Within this, we define trains, their routes, and components related to the train control system. Based on that, in Section~\ref{sec:tasks} we formalize the design tasks, which might arise in the context of ETCS~L2~HTD.

\subsection{Basic Definition of Railway Networks}
\begin{figure}[!t]
	\centering
	\subfloat[Full model as directed graph]{\includegraphics[width=0.45\textwidth]{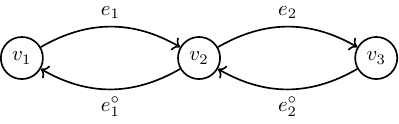}\label{fig:single_track_full}}
	\hfill
	\subfloat[Simplified visualization]{\includegraphics[width=0.45\textwidth]{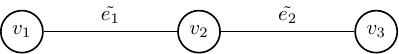}\label{fig:single_track_simplified}}
	\caption{Graphs representing a single track}
\end{figure}
\begin{figure}[!t]
	\centering
	\begin{minipage}{0.48\textwidth}
		\includegraphics[width=\textwidth]{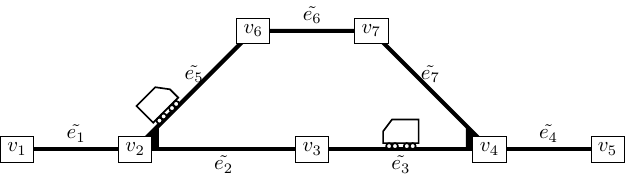}
		\caption{Siding as a graph}
		\label{fig:track_with_siding}
	\end{minipage}\hfill
	\begin{minipage}{0.48\textwidth}
		\includegraphics[width=\textwidth]{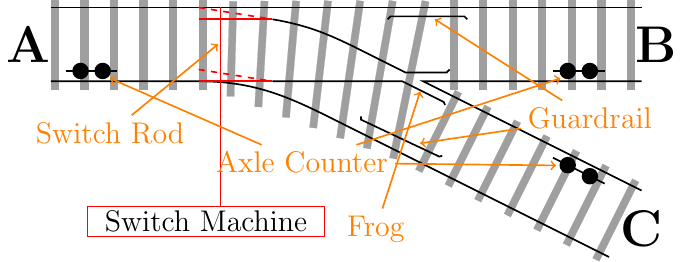}
		\caption{Simple turnout}
		\label{fig:turn_out}
	\end{minipage}
\end{figure}

In the following, we derive a definition for a railway network and valid movements on it that provide the basis for formulating corresponding design and verification tasks. For this, we introduce a directed graph \mbox{$G=(V,E)$} together with further properties and constraints. In a straightforward and abstract view, edges~$E$ represent tracks that connect to the nodes~$V$ representing certain \emph{points of interests} of the network (e.g., turnouts, crossings, signals, and stations). While railway tracks do not have a direction, a notion of direction will be important when introducing trains to prevent corner cases that are cumbersome to exclude in an undirected setting. Unless special operational constraints limit the direction on a track, we add edges for both directions; see Figure~\ref{fig:single_track_full} representing a single track on which trains can move from left to right and from right to left. However, as said, edges $(v_1,v_2)$ and $(v_2,v_1)$, for example, represent the same track segment. For edge $e=(u,v) \in E$, we denote its reverse edge $e^\circ=(v,u)$ if existent. In particular, $e$ and $e^\circ$ represent the same track segment and $\left(e^\circ\right)^\circ=e$.
\begin{remark}
	To simplify notation and visualization, we let $\tilde{e}$ denote both $e$ and $e^\circ$ and visualize this by an undirected edge. Hence, the graph in Figure~\ref{fig:single_track_full} simplifies to Figure~\ref{fig:single_track_simplified}. Whenever we use an undirected edge $\tilde{e}$, it should be clear from the context to which of the two directed edges we refer. While this notation is rather imprecise, we believe it will improve the readability of some concepts as long as we bear in mind that they represent two edges simultaneously.
\end{remark}
Of course, an edge~$e\in E$ has a natural length, e.g., in meters, which we denote by a mapping $\len \colon  E \to \Rn$. Note that $\len(e)=\len(e^\circ)$ since they represent the same track segment.
\begin{example}\label{ex:track_with_siding}
	Fig.~\ref{fig:track_with_siding} shows a simple railway layout with a siding. Because of this, it is possible that two trains driving in opposite directions can pass each other. For example, the train driving from $v_1$ to $v_5$ can use the turnout at $v_2$ via $v_6$ and $v_7$ and come back to the main track using turnout $v_4$ and, then, drive towards $v_5$. The second train, driving from $v_5$ to $v_1$, can drive via $v_4$, $v_3$, and $v_2$ to its final destination $v_1$. In this example, nodes $v_2$ and $v_4$ represent simple turnouts.
\end{example}

Turnouts usually do not allow trains to arbitrarily switch between the conjoining tracks. Instead, the possible ways over the turnout depend (besides others), e.g., on the direction the train is coming from. More precisely, Figure~\ref{fig:turn_out} shows a schematic drawing of a simple turnout as a standard \mbox{right-hand} switch with all its details. A train coming from the \mbox{left-hand} side (marked with an \textbf{A}) can take two ways: either to \textbf{B} or \textbf{C}. In contrast, a train coming from the right-hand side (either from~\textbf{C} or \textbf{B}) can take only one way, namely to \textbf{A}. These restrictions on movements are not yet incorporated within the described graph framework. Hence, we introduce a successor function $s_v \colon \delta^{in}(v) \to \mathcal{P}\left(\delta^{out}(v)\right)$ for every vertex $v \in V$, where $\delta^{in}(v)$ and $\delta^{out}(v)$ are the sets of edges entering and leaving $v$ respectively. Then, $s_v(e_{in})$ represents the possible successor edges when entering $v$ from a certain edge $e_{in} \in \delta^{in}(v)$.
\begin{example}
	In the case of Example~\ref{ex:track_with_siding}, the successor functions $\{s_v\}_{v \in V}$ are defined by
	\begin{align}
		s_{v_3}(e) &= \begin{cases}
			\{\tilde{e_3}\}, & \text{if } e=\tilde{e_2} \\
			\{\tilde{e_2}\}, & \text{if } e=\tilde{e_3}
		\end{cases} &%
		s_{v_7}(e) &= \begin{cases}
			\{\tilde{e_7}\}, & \text{if } e=\tilde{e_6} \\
			\{\tilde{e_6}\}, & \text{if } e=\tilde{e_7}
		\end{cases} \\
		s_{v_6}(e) &= \begin{cases}
			\{\tilde{e_6}\}, & \text{if } e=\tilde{e_5} \\
			\{\tilde{e_5}\}, & \text{if } e=\tilde{e_6}
		\end{cases} &%
		s_{v_1}(\tilde{e_1}) &= s_{v_5}(\tilde{e_4}) = \emptyset
	\end{align}
	for the nodes corresponding to straight tracks without any turnouts and
	\begin{align}
		s_{v_2}(e) & = \begin{cases}
			\{\tilde{e_2},\tilde{e_5}\}, & \text{ if } e = \tilde{e_1} \\
			\{\tilde{e_1}\}, & \text{ if } e \in \{\tilde{e_2},\tilde{e_5}\}
		\end{cases} &%
		s_{v_4}(e) & = \begin{cases}
			\{\tilde{e_3},\tilde{e_7}\}, & \text{ if } e = \tilde{e_4} \\
			\{\tilde{e_4}\}, & \text{ if } e \in \{\tilde{e_3},\tilde{e_7}\}
		\end{cases}
	\end{align}
	for the functions corresponding to turnouts.
\end{example}

In practice, a considered railway network will be connected to neighboring networks that are out of scope for the respective design tasks.
Hence, trains will enter and exit the network under consideration at these borders.
\begin{definition}[Border Vertices]
	The set of \emph{border vertices} is denoted by $\mathcal{B} \subseteq V$ and all $v \in \mathcal{B}$ fulfill $|\Gamma(v)| = 1$, where $\Gamma(\cdot)$ denotes the neighboring vertices, i.e., they cannot be in the interior of the network.
	It is assumed that trains can only enter and exit the network with predefined headway times $h \colon \mathcal{B} \to \R_{\geq 0}$.
\end{definition}
\begin{example}
	In the case of Example~\ref{ex:track_with_siding}, $\mathcal{B} = \{v_1,v_5\}$ could hold with $h(v_1)=h(v_5)=120s$.
\end{example}
Using the above, we can now define a simple railway network as follows:
\begin{definition}[Railway Network]\label{def:railway_network}%
	A \emph{railway network} \mbox{$N=\left(G,\len,\left\{s_v\right\}_{v \in V}\right)$} is defined by
	\begin{itemize}
		\item a directed graph $G=(V,E)$ with vertices $V$ being the set of points of interest and edges $E$ being the railway tracks between the aforementioned points of interest,
		\item a mapping $\len \colon E \to \Rn$ denoting the length of each edge such that \mbox{$\len(e)=\len(e^\circ)$} for every pair of edges $e,e^\circ \in E$,
		\item a family of mappings $\left\{s_v\right\}_{v \in V}$, where \mbox{$s_v \colon \delta^{in}(v) \to \mathcal{P}\left(\delta^{out}(v)\right)$} represents the valid movements over $v$, and
		\item border vertices $\mathcal{B} \subseteq V$ together with headway times $h \colon \mathcal{B} \to \R_{\geq 0}$.
	\end{itemize}
\end{definition}

\begin{figure}[!t]
	\centering
	\begin{minipage}{0.3\textwidth}
		\includegraphics[width=\textwidth, angle = 90]{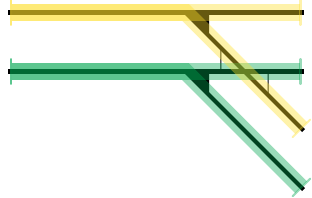}
		\caption{Double junction, or level junction}
		\label{fig:double_junction}
	\end{minipage}\hfill
	\begin{minipage}{0.23\textwidth}
		\includegraphics[width=\textwidth]{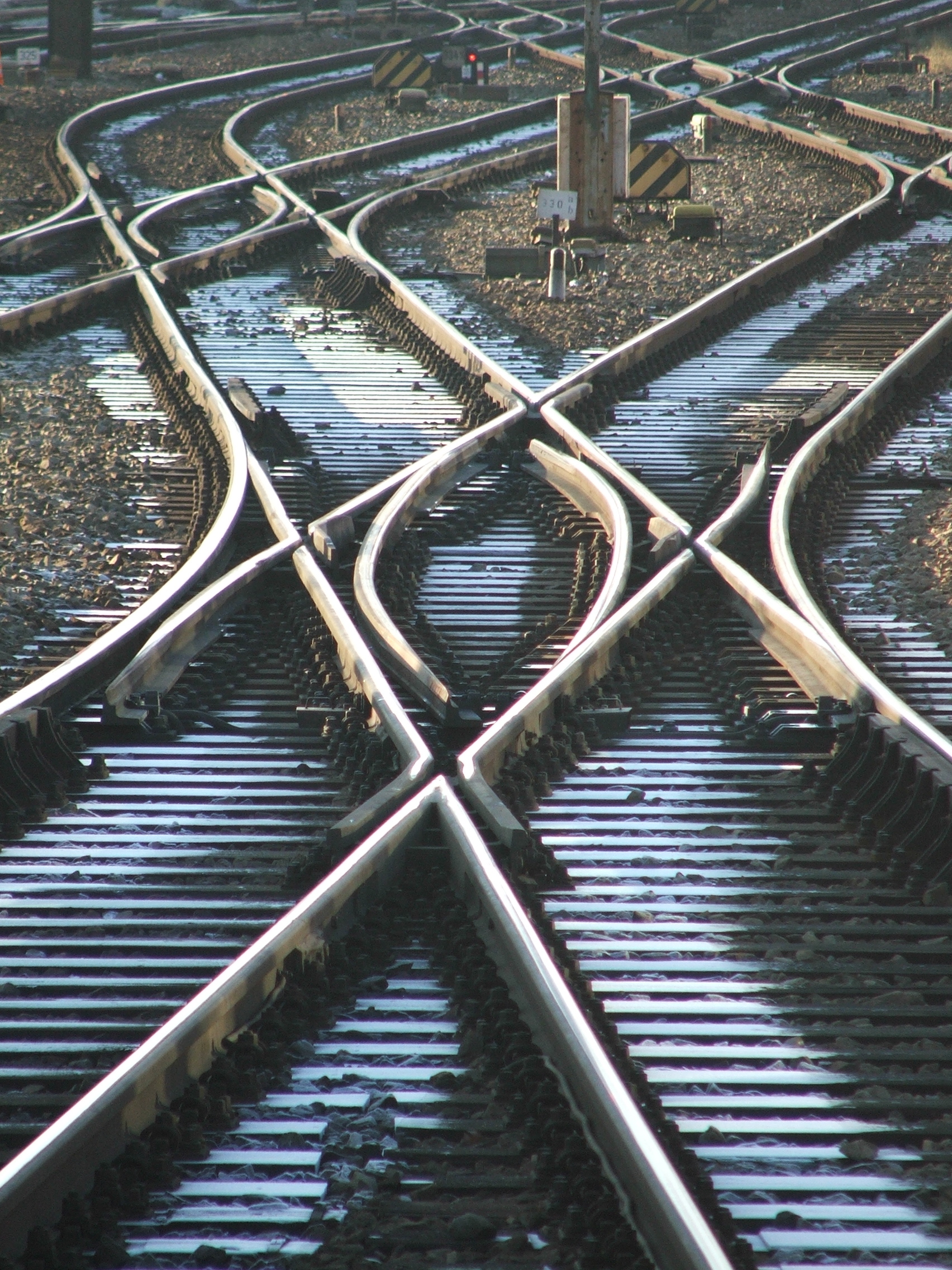}
		\caption{Double slip switch at Munich Central Station by Láczay \cite{Laczay2005} published under a CC BY-SA 2.0 license}
		\label{fig:dkw}
	\end{minipage}\hfill
	\begin{minipage}{0.4\textwidth}
		\includegraphics[width=\textwidth]{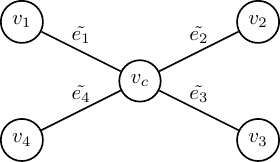}
		\caption{Graph of crossing}
		\label{fig:crossing_graph2}
	\end{minipage}
\end{figure}

However, railway networks do not only consist of turnouts where just three tracks are connected. For example, they might also cross each other. Reasons for this might be double junctions, where simple \emph{crossings} occur, as depicted in Figure~\ref{fig:double_junction} or \emph{slip switches} as depicted in Figure~\ref {fig:dkw}. The underlying graph structure is the same; however, the respective successor functions differ. In general, tracks that cross each other can be modeled as shown in Figure~\ref{fig:crossing_graph2}, where node $v_c$ corresponds to the crossing point.

In the case of a simple \emph{crossing}, e.g., where the green and yellow lines cross each other in Figure~\ref{fig:double_junction}, a train can move from $v_1$ ($v_2$) via $v_{c}$ to $v_{3}$ ($v_{4}$) or vice versa, but it is not possible that a train moves from $v_{1}$ via $v_{c}$ to $v_{2}$ or $v_{4}$. To model this movement restriction, the corresponding successor function $s_{v_c}$ is given by
\begin{align}
	s_{v_c}(\tilde{e_1}) & = \{\tilde{e_3}\}, & s_{v_c}(\tilde{e_2}) & = \{\tilde{e_4}\}, \\ s_{v_c}(\tilde{e_3}) & = \{\tilde{e_1}\} \qquad \text{and} & s_{v_c}(\tilde{e_4}) & = \{\tilde{e_2}\}.
\end{align}
On the other hand, \emph{double slip switches} as, e.g., shown in Fig.~\ref{fig:dkw} are quite common in the entry to bigger stations as their usage saves place, which is normally very limited in such.  Their switch machines make changing the track from one side to the other possible. In this case, the successor function at $v_c$ changes to
\begin{equation}
	s_{v_c}(e) = \begin{cases}
		\{\tilde{e_2},\tilde{e_3}\}, & \text{ if } e \in \{\tilde{e_1},\tilde{e_4}\} \\
		\{\tilde{e_1},\tilde{e_4}\}, & \text{ if } e \in \{\tilde{e_2},\tilde{e_3}\}
	\end{cases}.
\end{equation}
Likewise, a \emph{single slip switch} looks similar to double slip switches but allows fewer movements, i.e., for 2 out of 4 entry points, there is no possibility to change the track. More precisely, we can model this using
\begin{align}
	s_{v_c}(\tilde{e_1}) & = \{\tilde{e_2},\tilde{e_3}\}, & s_{v_c}(\tilde{e_3}) & = \{\tilde{e_1}\}, \\
	s_{v_c}(\tilde{e_2}) & = \{\tilde{e_1},\tilde{e_4}\} \qquad \text{and} & s_{v_c}(\tilde{e_4}) & = \{\tilde{e_2}\}.
\end{align}
These examples demonstrate that any real-world railway network can be modeled using Definition~\ref{def:railway_network}. Even if further not-mentioned points of interest occur, they can be represented analog to the above examples.

Note that the above definition is somewhat similar to the double-vertex graph introduced by Montigel \cite{Montigel1992} even though the successor functions are included in the double-vertices. At the same time, the choice to use these successor functions is similar to how valid train movements are modeled in \cite{Dang2019}. Because of how their specific MIP model is constructed, they do not require directed edges, which we believe is vital in our solution-independent description. Hence, the necessity to define train schedules, valid movements, and the possibility to add new sections differs from their work.

In principle, it is not important which of the above infrastructure models is used. We chose successor functions because we find this graph model to be more intuitive for the purpose of this work. The upcoming design tasks could easily also be formulated on, e.g., Montigel's double-vertex graph model\cite{Montigel1992}.

Using the above definition, we can also formally define valid track sequences, which could be traversed by a train and, hence, establish a connection between the edges of the graph and its successor functions.
\begin{definition}[Track Sequence]\label{def:track_sequence}
	For some $k \in \mathbb{N}$, a \emph{track sequence} is a $k$-tuple $(e_1, \ldots, e_k) \in E^k$ ($e_i = (v_i,w_i)$) such that $v_{i+1}=w_i$ for all $i=1,\ldots,k-1$. The track sequence is \emph{valid} if, and only if, $e_{i+1} \in s_{w_i}(e_i)$  for all $i=1,\ldots,k-1$.
\end{definition}

\begin{example}
	Consider again the network depicted in Figure~\ref{fig:track_with_siding}. Assume that all edges $e_i$ are facing right and $e_i^\circ$ are facing left. Then, $(e_2^\circ,e_5,e_6)$ is a track sequence since they form a path in the graph. However, the movement from $e_2^\circ$ to $e_5$ is forbidden by the successor function describing the logic of turnout $v_2$. Hence, the track sequence is not valid. On the other hand, $(e_1,e_5,e_6)$ is a valid track sequence.
\end{example}

Finally, observe that the above description does not model reality up to every detail. For example, the defined model cannot currently model dead-end stations. For this, trains must be allowed to turn around on certain edges after a predefined waiting time. However, we believe we have covered the most essential properties described above. Hence, it is reasonable to develop further work on the presented formulation first and, if they turn out promising, add further constraints afterward.

\subsection{Trains and Their Schedules}
Until now, we have focused on modeling the railway network without considering the trains and their movements. In this section, we define trains as well as their movements and schedules.
\begin{definition}[Train]
	A \emph{train} $\tr:=\left(l^{(\tr)},v_{max}^{(\tr)}, a^{(\tr)}, d^{(\tr)}, \tim^{(\tr)}\right)$ is a tuple, where
	\begin{itemize}
		\item $l^{(\tr)} \in \mathbb{R}_{> 0}$ defines its length and,
		\item $v_{max}^{(\tr)}$ defines its maximal possible speed,
		\item $a^{(\tr)}$ and $d^{(\tr)}$ denote the maximal acceleration and deceleration respectively, and
		\item $\tim^{(\tr)} \in \{true,false\}$ denotes if $\tr$ is equipped with \emph{train integrity monitoring}~(TIM).
	\end{itemize}
\end{definition}
Train lengths do not necessarily coincide with the length of edges within a railway network  \mbox{$N=\left(G,\len,\left\{s_v\right\}_{v \in V}\right)$}. In particular, trains might only be located on parts of an edge. For this, we denote a \emph{track interval} \mbox{$\iv := (e,\lambda, \mu) \in E \times [0,1] \times [0,1]$} such that $\lambda \leq \mu$, where~$\lambda$ denotes the start and $\mu$ the end of the section in relative terms.
\begin{definition}[Track Range]
	A \emph{track range}  \mbox{$\rg:=(s_{in}, \iv_1,\ldots,\iv_k, s_{out})$} is then given by a series of track intervals $\iv_i = (e_i,\lambda_i,\mu_i)$ together with $s_{in}, s_{out} \geq 0$. In this context, $s_{in}$ and $s_{out}$ denote the parts of the train that are outside of the network at the rear (incoming vertex) and front (outgoing vertex), respectively. The following conditions must hold by definition:
	\begin{enumerate}
		\item \mbox{$(e_1,\ldots,e_k)$} is a valid track sequence according to Definition~\ref{def:track_sequence},
		\item $\lambda_i = 0$ for every $1 < i \leq k$,
		\item $\mu_i = 1$ for every $1 \leq i < k$,
		\item $s_{in} \neq 0 \Rightarrow e_1 = (u,v)$ for some $u \in \mathcal{B}$ and $\lambda_1 = 0$, and,
		\item $s_{out} \neq 0 \Rightarrow e_k = (u,v)$ for some $v \in \mathcal{B}$ and $\mu_k = 1$.
	\end{enumerate}
\end{definition}
Naturally, the lengths of the edges carry over to track intervals and ranges by setting \mbox{$\len(\iv):=(\mu-\lambda) \cdot \len(e)$} and
\begin{equation}
	\len(\rg) := s_{in} + s_{out} + \sum_{i=1}^k \len(\iv_i) = s_{in} + s_{out} + \sum_{i=1}^k (\mu_i-\lambda_i) \cdot \len(e_i).
\end{equation}
If $k \geq 2$, then we can also write
\begin{equation}
	\len(\rg) = s_{in} + (1-\lambda_1)\cdot \len(e_1) + \sum_{i=2}^{k-1} \len(e_i) + \mu_k \cdot \len(e_k) + s_{out}.
\end{equation}

At this point, it would be intuitive to define the position of a train at a given time as a track range $\rg$ with $\len(\rg) = l^{(\tr)}$.
However, it is convenient to use a slightly different definition, which later makes it easier to properly define separation trains (or headways) of different trains with respect to each other.
Note that the entire distance needed to brake has to be already cleared from other trains.
Hence, one could say that a train does not only occupy the track parts on which it is physically present, but also all of its braking distance ahead.

\begin{definition}[Train Route and Occupation]
	For a given train $\tr$, a \emph{train route} $R^{(\tr)}_{[a,b]} := \left(\rg_t^{(\tr)}\right)_{t \in [a,b]}$ is a series of track ranges for every time $t \in [a,b]$, such that $\len\left(\rg_t^{(\tr)}\right) = l^{(\tr)} + \bd_t^{(\tr)}$, where $\bd_t^{(\tr)}$ denotes the distance $\tr$ requires to come to a full stop if it were to start decelerating at time $t$.
	In this context $\rg_t^{(\tr)}$ is bound by the trains rear end on the one and its moving authority on the other side. \label{def:train-route-occupation}
\end{definition}
Given some track range \mbox{$\rg:=\left(s_{in}, (e_1,\lambda_1,\mu_1),\ldots,(e_k,\lambda_k,\mu_k), s_{out}\right)$}, we denote
\begin{equation}
	\occ(\rg) := \{e_1,\ldots,e_k\}
\end{equation}
to be the occupied track segments. Naturally
\begin{equation}
	\occ_t^{(\tr)} := \occ\left(\rg_t^{(\tr)}\right)
\end{equation}
corresponds to the segments occupied by train $tr$ at time $t \in [a,b]$.

As of now, trains can theoretically move and jump around a railway network arbitrarily because the different time steps are not linked with each other. Consider two track ranges \mbox{$\rg^{(1)}=\left(s_{in}^{(1)},(e_1^{(1)},\lambda_1^{(1)},\mu_1^{(1)}),\ldots,(e_l^{(1)},\lambda_l^{(1)},\mu_l^{(1)}), s_{out}^{(1)}\right)$} and \mbox{$\rg^{(2)}=\left(s_{in}^{(2)}, (e_1^{(2)},\lambda_1^{(2)},\mu_1^{(2)}),\ldots,(e_k^{(2)},\lambda_k^{(2)},\mu_k^{(2)}), s_{out}^{(2)}\right)$}. They are said to \emph{intersect (in order)} if there exists an \mbox{$s \in \{1,2,\ldots,\min\{l,k\}\}$} such that \mbox{$e_{l-s+i}^{(1)}=e_i^{(2)}$} for all \mbox{$i \in \{1,\ldots,s\}$}, i.e., the end of $\rg^{(1)}$ and start of $\rg^{(2)}$ overlap. We then define
\begin{equation}
	\dist\left(\rg^{(1)},\rg^{(2)}\right) := \underbrace{\left(s_{in}^{(1)} - s_{in}^{(2)} \right)}_{\text{distance outside network}} + \overbrace{\sum_{i=1}^{l-s} \left(\mu_i^{(1)}-\lambda_i^{(1)}\right) \len\left(e_i^{(1)}\right)}^{\text{length of non-overlapping edges of } \rg^{(1)}} + \underbrace{\left(\lambda_1^{(2)}-\lambda_{l-s+1}^{(1)}\right) \len\left(e_{l-s+1}^{(1)}\right)}_{\text{length on overlapping edge}}
\end{equation}
to be their \emph{distance}. Using this, we can characterize the varying speed of a train at time $t$ by the following definition.
\begin{definition}[Train Speed]\label{def:speed}
	A train \mbox{$tr:=\left(l^{(\tr)},v_{max}^{(\tr)},\left(\rg_t^{(\tr)}\right)_{t\geq 0}\right)$} is said to \emph{respect its maximal speed} if for every $t \geq 0$ and $0<\varepsilon < \frac{l^{(\tr)}}{v_{max}^{(\tr)}}$ it holds that $\rg_t^{(\tr)}$ and $\rg_{t+\varepsilon}^{(\tr)}$ intersect in order with
	\begin{equation}
		\dist\left(\rg_t^{(\tr)},\rg_{t+\varepsilon}^{(\tr)}\right) \leq \varepsilon \cdot v_{max}^{(\tr)}.
	\end{equation}
	Its \emph{speed} at time $t$ is then given by
	\begin{equation}
		v_t^{(\tr)} := \frac{\text{d}}{\text{d}\varepsilon}\left[\dist\left(\rg_t^{(\tr)},\rg_{t+\varepsilon}^{(\tr)}\right)\right]_{\varepsilon=0} = \lim_{\varepsilon \to 0}\frac{\dist\left(\rg_t^{(\tr)},\rg_{t+\varepsilon}^{(\tr)}\right)}{\varepsilon}.
	\end{equation}
	Moreover, the change in speed is bound by
	\begin{equation}
		-d^{(\tr)} \leq \frac{d}{dt} \left( v_t^{(\tr)} \right) \leq a^{(\tr)}.
		\label{eqn:a-d-bound}
	\end{equation}
	A train \emph{respects its braking distance} if for every $t \geq 0$
	\begin{equation}
		\bd_t^{(\tr)} \geq \frac{v_t^{(\tr)}}{2 \cdot d^{(\tr)}}. \label{eqn:braking-distance}
	\end{equation}
\end{definition}
By definition, $t \in \mathbb{R}$, so that train velocity and occupation are defined continuously.
When solving design tasks proposed in Section~\ref{sec:tasks}, time might or might not have to be discretized depending on the applied algorithm.
However, this does not affect the problem description itself, which assumes continuous time and is independent of proposed solving strategies.

\begin{figure}[!t]
	\centering
	\begin{minipage}[b]{0.42\textwidth}
		\includegraphics[width=\textwidth]{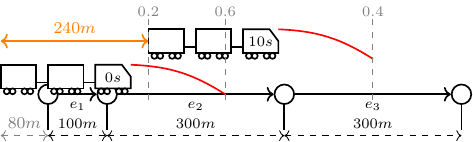}
		\caption{A train $tr$ at two overlapping time steps $t$ and $t+\varepsilon$}
		\label{fig:train-speed}
	\end{minipage}\hfill
	\begin{minipage}[b]{0.42\textwidth}
		\includegraphics[width=\textwidth]{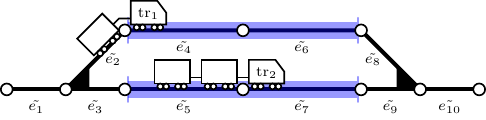}
		\caption{Example train station}
		\label{fig:station}
	\end{minipage}\hfill
	\begin{minipage}[b]{0.14\textwidth}
		\includegraphics[width=\textwidth]{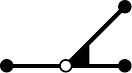}
		\caption{Turnout}
		\label{fig:turnout-as-graph}
	\end{minipage}
\end{figure}

Of course, further details of train movements can be modeled within this framework as additional constraints.
For example, the train's acceleration and deceleration properties might depend on its current speed and/or the underlying track's gradient.
Hence, braking curves defined in the official specifications can also be considered \cite{EUAR2020} by altering Equations~(\ref{eqn:a-d-bound}) and~(\ref{eqn:braking-distance}) respectively.
Usually, trains should not face red signals when planning their movements.
Such a requirement could, e.g., be included as further constraints on the speed to disallow stopping on an open line.
\begin{example}
	Consider Figure~\ref{fig:train-speed} with an \mbox{210m-long} train occupying $(80,(e_1,0,1),(e_2,0,0.6),0)$, which moved to $(0,(e_2,0.2,1),(e_3,0,0.4))$ within $10$~seconds (we have $\bd_{0s}^{(\tr)}=\bd_{10s}^{(\tr)}=150m$). The distance $\dist\left(\rg_{0s}^{(\tr)},\rg_{10s}^{(\tr)}\right)$ is given by \mbox{$\underbrace{(80m-0m)}_{s_{in}^{(0s)} - s_{in}^{(10s)}} + \underbrace{(1-0)}_{\mu_1^{(0s)}-\lambda_1^{(0s)}} \cdot \underbrace{100m}_{\len(e_1)} + \underbrace{(0.2-0)}_{\lambda_1^{(10s)}-\lambda_2^{(0s)}} \cdot \underbrace{300m}_{\len(e_2)} = 240m$}. Hence, we can approximate its speed by $\frac{240m}{10s}=24\frac{m}{s}$.
\end{example}

Of course, a train does not drive anywhere on a network but has a fixed set of stations to visit or possibly even a fixed schedule. We denote a \emph{train station} to be a subset of edges $S \subseteq E$. A train $tr$ is said to be \emph{in} station $S$ at time $t$ if $\occ_t^{(\tr)} \subseteq S$. It \emph{stops} in $S$ if additionally $v_t^{(\tr)} = 0$. For example, a station can include all tracks of a certain \mbox{real-world} train station. If necessary, we can also restrict it to only cover a subset, e.g., if some platforms should be used solely by regional and others only by long-distance trains.
\begin{example}
	Consider Figure~\ref{fig:station}, a station \mbox{$S=\{\tilde{e_4},\tilde{e_5},\tilde{e_6},\tilde{e_7}\}$}, and the two trains depicted. Here, $\tr_1$ occupies $\tilde{e_2}$ and $\tilde{e_4}$. Since \mbox{$\tilde{e_2}\not\in S$}, $\tr_1$ is not in the train station. On the other hand, $\tr_2$ occupies \mbox{$\tilde{e_5} \in S$} and \mbox{$\tilde{e_7} \in S$} and, hence, is in the train station.
\end{example}
Combining the above and adding time constraints, we are now ready to define train schedules.
\begin{definition}[Timetable Requests]\label{def:schedule}
	Let $tr_1,\ldots,tr_m$ be $m$ trains on a railway network $N$. For every train $tr_i$, a \emph{timetable request} consists of
	\begin{itemize}
		\item an entry node $v_{in}^{(tr_i)}$ together with a desired arrival interval $[\tl_{in}^{(tr_i)},\tu_{in}^{(tr_i)}]$,
		\item an exit node $v_{out}^{(tr_i)}$ together with a desired departure interval $[\tl_{out}^{(tr_i)},\tu_{out}^{(tr_i)}]$, as well as,
		\item a list of $m_i$ stations $S_1^{(tr_i)},\ldots,S_{m_i}^{(tr_i)}\subseteq E$ together with
		\begin{itemize}
			\item desired arrival intervals \mbox{$[\al_1^{(tr_i)},\au_1^{(tr_i)}],\ldots,[\al_{m_i}^{(tr_i)},\au_{m_i}^{(tr_i)}]$},
			\item desired departure intervals \mbox{$[\dl_1^{(tr_i)},\du_1^{(tr_i)}],\ldots,[\dl_{m_i}^{(tr_i)},\du_{m_i}^{(tr_i)}]$}, and,
			\item minimal stopping times $\Delta t_1^{(tr_i)}, \ldots, \Delta t_{m_i}^{(tr_i)} \geq 0$.
		\end{itemize}
	\end{itemize}
	A train $tr_i$ \emph{respects its timetable request} if, 
	\begin{itemize}
		\item it enters the network at node $v_{in}^{(tr_i)}$ at some time $t_{in} \in [\tl_{in}^{(tr_i)},\tu_{in}^{(tr_i)}]$, i.e., $s_{in}^{t} = \len(tr_i)$ and $e_{in} \in \occ_{t+dt}^{(tr_i)}$ for all $0 < dt < \varepsilon$ for some $\varepsilon > 0$, where $e_{in} \in \delta^{out}\left(v_{in}^{(tr_i)}\right)$,
		\item  it leaves the network at node $v_{out}^{(tr_i)}$ at some time $t_{out} \in [\tl_{out}^{(tr_i)},\tu_{out}^{(tr_i)}]$, i.e., $s_{out}^{t} = \len(tr_i)$ and $e_{out} \in \occ_{t-dt}^{(tr_i)}$ for all $0 < dt < \varepsilon$ for some $\varepsilon > 0$, where $e_{out} \in \delta^{in}\left(v_{out}^{(tr_i)}\right)$, and,
		\item for every \mbox{$j \in \{1,\ldots,m_i\}$}, there exists an interval $[\tl_j^{(tr_i)},\tu_j^{(tr_i)}]$ with $\tu_j^{(tr_i)} - \tl_j^{(tr_i)} \geq \Delta t_j^{(tr_i)}$, $\tl_j^{(tr_i)} \in [\al_j^{(tr_i)},\au_j^{(tr_i)}]$, and $\tu_j^{(tr_i)} \in [\dl_j^{(tr_i)},\du_j^{(tr_i)}]$, such that $tr_i$ stops in $S_{j}^{(tr_i)}$ for every $t \in [\tl_j^{(tr_i)},\tu_j^{(tr_i)}]$.
	\end{itemize}
\end{definition}

\subsection{Sections and Virtual Subsections in Railway Networks}\label{sec:TTD}
As mentioned, trains cannot drive independently from each other, and running into each other is an unwanted situation. For this, a control system ensures that a block is occupied by at most one train, which still needs to be defined formally. To this end, \emph{Trackside Train Detection}~(TTD) systems, e.g., axle counters, are used to collect corresponding information on the occupation of blocks. Accordingly, corresponding blocks are frequently also called TTD sections.

Unfortunately, an edge~$e \in E$ of the graph definition from above does not match a TTD section. A TTD section starts and ends with, e.g., axle counters. Three axle counters are usually used for a turnout: one for each of the three tracks that meet at the turnout (see also Figure~\ref{fig:turn_out}). In this case, the switch's three-track segments (edges) belong to the same block. To model such TTD sections, we introduce another helping function on the vertices, also called border function, \mbox{$C_V \colon V \to \{0=\text{no border}, 2=\text{TTD-Border}\}$}, which indicates if a certain vertex is equipped with trackside train detection and, thus, corresponds to a border between two TTD sections. A simple turnout would typically correspond to one central node of value~0 and three outer nodes of value~2, in which case the respective edges will have a rather small length corresponding to the distance between the turnout's center and the axle counters. This is depicted in Figure~\ref{fig:turnout-as-graph}: A black filled node \begin{tikzpicture} \draw[thick, fill=black] (0,0) circle (0.1); \end{tikzpicture} visualizes a vertex which represents a TTD-Border, i.\,e., $C_V$ evaluates to $2=\text{TTD-Border}$.
A white filled circle \begin{tikzpicture} \draw[thick, fill=white] (0,0) circle (0.1); \end{tikzpicture} visualizes a vertex which represents no TTD-Border, i.\,e., $C_V$ evaluates to $0=\text{no border}$.
Abstractly, we refer to a vertex $v \in V$ with $C_V(v) = k$ to be a border of level $k$. Further, we define by $L'_k(G):=(V'_k,E'_k)$ an \emph{undirected} graph interchanging the role of vertices and edges and additionally ignoring borders of level $\geq k$.

\begin{figure}[!t]
	\centering
	\includegraphics[width=0.98\textwidth]{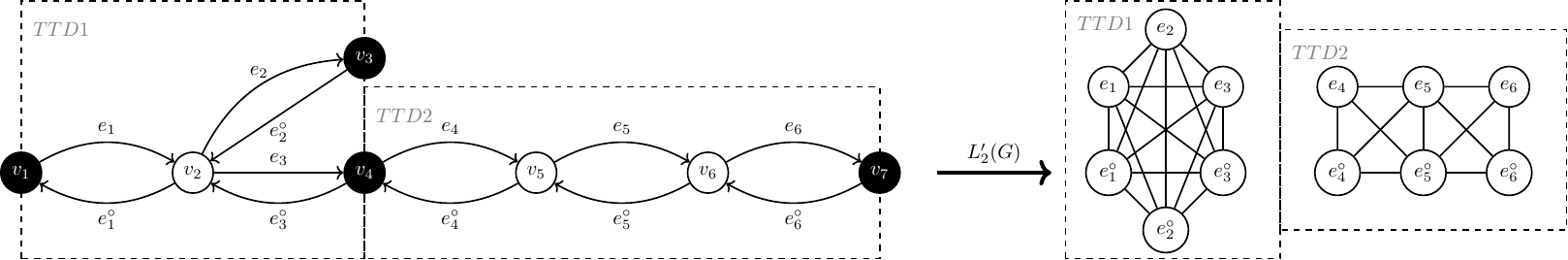}
	\caption{TTD sections}
	\label{fig:line-graph}
\end{figure}

Formally, we define $V'_k := E$ and
\begin{equation}
	E'_k := \left\{\{e_1,e_2\} \subseteq V'_k \colon e_2 = e_1^\circ \text{ or } \tilde{e_1} \cap \tilde{e_2} \cap C_K \neq \emptyset\right\},
\end{equation}
where $C_K := \{v \in V \colon C_V(v) < k\}$ are the nodes with border value less than $k$. Now, if two nodes within $L'_2(G)$ are connected, then their corresponding track segments are not separated by any TTD; hence, they fall within one TTD section. In particular, if we set
\begin{equation}
	TTDs := \{T_1,\ldots,T_m\} = CC\left(L'_2(G)\right),\label{eqn:TTDs}
\end{equation}
where $CC(\cdot)$ denotes the connected components, then $T_1,\ldots,T_m$ correspond exactly to the TTD sections. Edges within a section $T_i \in TTDs$ are not separated by any TTD, whereas a TTD section $T_i \in T$ is separated from any other TTD section $T_j \in TTDs-\{T_i\}$ by corresponding borders. Naturally, this provides a partition of the edges of the original graph $G$.
\begin{example}
	Consider a graph with seven vertices as depicted in Figure~\ref{fig:line-graph}.
	Vertices $v_1$ to $v_4$ represent a simple turnout as in Figure~\ref{fig:turn_out} connected to a straight track up to $v_7$.
	As before, \begin{tikzpicture} \draw[thick, fill=black] (0,0) circle (0.1); \end{tikzpicture} corresponds to a TTD border and \begin{tikzpicture} \draw[thick, fill=white] (0,0) circle (0.1); \end{tikzpicture} to no border.
	Figure~\ref{fig:line-graph} shows how the network transforms into $L'_2(G)$, which consists of two connected components.
	The vertices of each connected component (i.e., edges in the original graph) compose one TTD section.
	The corresponding sections are marked in both graphs in Figure~\ref{fig:line-graph}.
\end{example}

With the introduction of hybrid train detection, sections do not have to be separated by TTDs. In particular, this allows splitting existing TTD sections into virtual subsections (VSS). To also model these, we add a VSS-Border element to the border function as follows:
\mbox{$C_V \colon V \to \{0=\text{no border}, 1=\text{VSS-Border}, 2=\text{TTD-Border}\}$}.
In figures, we plot VSS borders using \begin{tikzpicture} \draw[preaction={fill, white}, pattern=north east lines, pattern color=black] (0,0) circle (0.1); \end{tikzpicture}.
Similarly to Equation~(\ref{eqn:TTDs}) the VSS sections $V_1,\ldots,V_n$ are given by the connected components of $L'_1(G)$, i.e.,
\begin{equation}
	\{V_1,\ldots,V_n\} := CC\left(L'_1(G)\right).\label{eqn:vss}
\end{equation}
A key concept in train control is the division of the tracks into multiple blocks. A block is only allowed to be occupied by at most one train. In the context of ETCS~L2~HTD, these blocks are usually given by the VSS sections that function essentially analog to their TTD counterparts.

\begin{figure}
	\centering
	\includegraphics[width=0.95\textwidth]{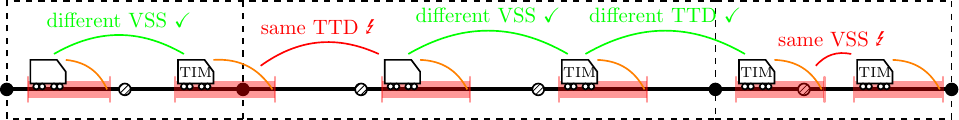}
	\caption{VSS condition}
	\label{fig:vss-condition}
\end{figure}

The concrete control logic depends on if a train is equipped with \emph{train integrity monitoring}~(TIM) or not.
A train with TIM can safely reports its rear end, because it is determined with sufficient certainty that the train is still complete and has not lost any wagons.
Because of this a VSS section can be safely reported as free without the need of any TTD hardware.
Hence, any other train is permitted to enter this VSS that has just been cleared.
On the other hand, a train without TIM can only safely report its front position.
Thus, a VSS remains occupied until the corresponding TTD section is cleared by the trackside hardware.
These cases are depicted in Figure~\ref{fig:vss-condition}, in which only 4 out of 6 trains are equipped with TIM.

Hence, the respective control condition can be formalized as follows:
\begin{definition}[VSS Condition]\label{def:vss-condition}
	Let $V_1,\ldots,V_n$ be the \mbox{VSS-sections} as defined in Equation~(\ref{eqn:vss}) and $tr_1,\ldots,tr_m$ be the trains operating on the respective railway network. The train control system ensures the following condition for every $i \in \{1,\ldots,n\}$, $j \in \{1,\ldots,m\}$, $k \in \{1,\ldots,m\}-\{j\}$, and $t \geq 0$:
	\begin{equation}
		V_i \cap \occ_t^{\left(tr_j\right)} \neq \emptyset \Rightarrow V_i \cap \occ_t^{\left(tr_k\right)} = \emptyset.\label{eqn:vss-constraint}
	\end{equation}
	Furthermore, let $T_1,\ldots,T_l$ be the \mbox{TTD-sections} as defined in  Equation~(\ref{eqn:TTDs}). For any $h \in \{1,\ldots,l\}$, $j \in \{1,\ldots,m\}$, and $t \geq 0$ with $T_h \cap \occ_t^{(\tr_j)} \neq \emptyset$ let
	\begin{equation}
		\hat{t}^{(\tr_j)}_h(t) := \max\left\lbrace 0, \sup\left\lbrace \tau \colon 0 \leq \tau \leq t, T_h \cap \occ_\tau^{(\tr_j)} = \emptyset \right\rbrace \right\rbrace.
	\end{equation}
	Using the above definition, for every $j \in \{1,\ldots,m\}$ with $\tim^{(\tr_j)}=false$ as well as $k \in \{1,\ldots,m\}-\{j\}$, $h \in \{1,\ldots,l\}$, $i \in \{1,\ldots,n\}$, $t \geq 0$ and $\tau \in \left[\hat{t}^{(\tr_j)}_h(t),t\right]$ it has to additionally hold that
	\begin{equation}
		T_h \cap V_i \cap \occ_\tau^{(\tr_j)} \neq \emptyset \Rightarrow V_i \cap \occ_t^{(\tr_k)} = \emptyset.\label{eqn:non-tim-constraint}
	\end{equation}
\end{definition}
Intuitively, in Definition~\ref{def:vss-condition}, Equation~(\ref{eqn:vss-constraint}) ensures that only one VSS is occupied at any given time.
$\hat{t}^{(\tr_j)}_h(t)$ is the time at which $\tr_j$ entered (began to occupy) $T_h$. 
Equation~(\ref{eqn:non-tim-constraint}) simply states that if a train without TIM occupies a TTD section $T_h$ any VSS within $T_h$ previously occupied by that train is not yet cleared.

Because the braking distance was already included in each trains occupation (see Definition~\ref{def:train-route-occupation}), the conditions in Definition~\ref{def:vss-condition} ensure the safe distance required between following trains.

\begin{remark}
	If all trains are equipped with TIM, only Equation~(\ref{eqn:vss-constraint}) remains in Definition~\ref{def:vss-condition}.
	Using an ETCS L2 control with fixed TTD sections and no VSS, Definition~\ref{def:vss-condition}, again, simplifies to Equation~(\ref{eqn:vss-constraint}) with $T_1,\ldots,T_h$~(TTD-sections) instead of $V_1,\ldots,V_n$~(VSS-sections), i.e., $T_i \cap \occ_t^{\left(tr_j\right)} \neq \emptyset \Rightarrow T_i \cap \occ_t^{\left(tr_k\right)} = \emptyset$.
\end{remark}

Overall, given a railway network $N$, a certain number of trains, and a list of timetable requests, the following properties are desired whenever considering a design task:
\begin{itemize}
	\item Every train respects its maximal speed and braking distance (Definition~\ref{def:speed})
	\item Every train respects its timetable request (Definition~\ref{def:schedule}).
	\item The VSS condition is satisfied (Definition~\ref{def:vss-condition}).
\end{itemize}

Again, note that one can, in theory, model the behavior of the control system in as much detail as desired.
For example, interlockings or time needed for turnouts to change their position could be considered using additional constraints.
Nevertheless, we believe the level of detail provided in this paper suffices for defining relevant design tasks.

\section{Design Tasks and their Complexity}\label{sec:tasks}
In this section, we define different design tasks that might occur in ETCS~L2~HTD systems and could be valuable to automate. They are formulated independently from whichever solution method might be chosen to solve them.
In Section~\ref{sec:ver}, we formulate a respective verification tasks.
While the verification is arguably more of a theoretical nature, it provides the basis for the design tasks proposed in Sections~\ref{sec:vss_layouts} and~\ref{sec:schedule}.
Especially the optimization tasks in Section~\ref{sec:schedule}, which use the added flexibility by HTD control systems to use specific timetable requests instead of general schedule-independent KPIs, are both, of practical interest and not yet well studied.

The proposed design tasks in the following subsections are formulated to help infrastructure planners to find optimal signaling layouts in terms of enabling the best operational outcome based on specific timetable requests.
They do not yet consider robustness with respect to delays.
For classical signaling systems capacity was already considered under disturbed conditions leading to real-time rescheduling problems \cite{Goverde2013}.
Recently such rescheduling problems have also been considered for modern (moving block) control systems \cite{Versluis2024}.
At the same time, previous work focuses on minimizing the effect of delays on a fixed signaling layout.
An interesting extension to the presented design tasks would be to also consider stochastic distributions on expected delays.
In that case one could aim to optimize the VSS layout to minimize the negative effects of expected rescheduling during operation.
At the time of writing, the proposed design tasks are not well studied even in the deterministic case.
Hence, for now, we focus on providing efficient solutions for the proposed design tasks.
Once reasonable methods exist, adding expected stochastic delays would be a natural extension. 

\subsection{Verification of Timetable Requests on ETCS~L2~HTD Layouts}\label{sec:ver}
\begin{figure*}[!t]
	\centering
	\includegraphics[width=0.7\textwidth]{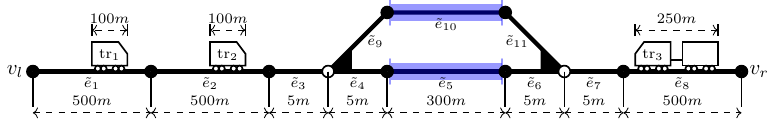}
	\caption{Infeasible verification task}
	\label{fig:infeasible-verification}
\end{figure*}

Note that a desired timetable request might be impossible on a particular railway network since only one train can be located on each TTD/VSS section, a restricted track speed, or a limited train speed. 
\newpage
\begin{problem}
	\problemtitle{Verification of Timetable Requests (TR-VER)}
	\probleminput{A railway network $N=(G,\len,\{s_v\}_{v \in V})$; border function $C_V \colon V \to \{0,1,2\}$; trains $\tr_1,\ldots,\tr_m$; a list of timetable requests for $\tr_1$ to $\tr_m$.}
	\problemquestion{Do there exist train routes $R^{(\tr_1)},\ldots,R^{(\tr_m)}$ such that the VSS condition is fulfilled and all trains respect their speed and timetable requests? If yes, provide a certificate.}
\end{problem}
\begin{example}
	Consider the train network in Figure~\ref{fig:infeasible-verification} and three trains $\tr_1$ to $\tr_3$. As before, \begin{tikzpicture} \draw[thick, fill=black] (0,0) circle (0.1); \end{tikzpicture} corresponds to a TTD border and \begin{tikzpicture} \draw[thick, fill=white] (0,0) circle (0.1); \end{tikzpicture} to no border. Also, we are given one real-world train station $S := \{\tilde{e_{5}},\tilde{e_{10}}\}$, which is marked in blue. The timetable requests are specified as follows:
	\begin{itemize}
		\item[$\tr_1$:] Entering at $v_l$ at $t=120$ (i.e., within [120,120]), leaving at $v_r$ at $t = 645$ (i.e., within [645, 645]), and stopping in $S$ from $t=240$ to $t = 300$ (i.e., arriving within [120,240], leaving within [300,645], and stopping for a minimal time of $\Delta t = 60$).
		\item[$\tr_2$:] Entering at $v_l$ at $t = 0$, leaving at $v_r$ at $t = 420$, and stopping in $S$ from $t=120$ to $t = 300$.
		\item[$\tr_3$:] Entering at $v_r$ at $t = 0$, leaving at $v_l$ at $t = 420$, and stopping in $S$ from $t = 180$ to $t = 300$.
	\end{itemize}
	where time is measured in seconds. Observe that the station $S$ only consists of two blocks; hence, a maximum of two trains can stop at a time. However, the timetable requests require all three trains to be stopped at the station at time $t=270$, for example. Hence, there are no train routes such that the VSS condition is fulfilled, the trains respect their timetable requests, and the verification task returns infeasibility.
	\label{ex:verification}
\end{example}

Observe that the routing on a fixed VSS layout might differ slightly from systems using trackside-oriented train detection, see also discussion around Definition~\ref{def:vss-condition}.
This is because trains equipped with \emph{train integrity monitoring}~(TIM) can report VSS sections free independently of their corresponding TTD sections.
Any train can follow closely.
At the same time, trains without TIM need TTD hardware (e.g., axle counters) to report a section as free.
Hence, a train can only follow if a non-TIM train has cleared an entire TTD section.

If all trains are equipped with TIM, the situation simplifies.
In that case, there is no logical difference between TTD and VSS sections, and the block layout is fixed for the verification tasks defined above.
Hence, TR-VER is essentially equivalent to the task on classical layouts in this (simplified) case.
Scheduling under such classical signaling systems is known to be NP-hard in general \cite{Mascis2002, DAriano2007}.
In particular, it is reasonable to assume that \mbox{TR-VER} has similar complexity.
At the same time, there are minor differences to the scheduling problems considered so that the proof does not carry over formally.
While the following complexity result is not unexpected, we formally include them for reasons of completeness.
Moreover, it serves as a basis for the following design tasks in Sections~\ref{sec:vss_layouts}~and~\ref{sec:schedule}, which are tailored to ETCS~L2~HTD.

\begin{theorem}
	TR-VER is NP-complete even if all routes and exact stopping times in train stations are fixed and the railway network is planar. \label{thm:complexity-verification}
\end{theorem}
\begin{proof}
	Given a certificate, we can check feasibility using the definitions in polynomial time; hence, TR-VER is in NP. 
	To show NP-completeness, we reduce Monotone 3-SAT-($\leq 2, \leq 2$) to TR-VER. Monotone 3-SAT-($\leq 2, \leq 2$) are all SAT instances in which every clause contains three literals (all of which contain only unnegated or only negated variables), and every variable appears in clauses at most twice unnegated and twice negated.
	Monotone 3-SAT-($\leq 2, \leq 2$) is known to be NP-complete \cite{Darmann2021}.
	
	For reduction, we are given an instance of Monotone 3-SAT-($\leq 2, \leq 2$) on variables $x_1, \ldots, x_n$ defined by clauses $\mathcal{C} = \{C_1, \ldots, C_m\}$, each consisting of three literals.
	For every clause, we add two vertices $c_{in}^{(j)}$ and $c_{out}^{(j)}$ for $j = 1,\ldots,m$. 
	Moreover, we add nodes $x_0$ and $x_{n+1}$.
	W.l.o.g., assume that clauses $C_1, \ldots, C_k$ contain only nonnegated literals and clauses $C_{k+1},\ldots,C_m$ contain only negated literals.
	Then we add edges $(c_{in}^{(j)}, x_0)$ and $(x_{n+1}, c_{out}^{(j)})$ for every $1 \leq j \leq k$ as well as $(c_{in}^{(j)}, x_{n+1})$ and $(x_0, c_{out}^{(j)})$ for every $l+1 \leq j \leq m$.
	For every literal, we add vertices $x_j^{(1)}, x_j^{(2l)}, x_j^{(2)}, x_j^{(2r)},$ and $x_j^{(3)}$ for every $1 \leq j \leq n$.
	We connect them using edges $(x_j^{(1)}, x_j^{(2r)})$, $(x_j^{(2r)}, x_j^{(3)})$, $(x_j^{(3)}, x_j^{(2l)})$, $(x_j^{(2l)}, x_j^{(1)})$, $(x_j^{(1)}, x_j^{(2)})$, $(x_j^{(2)}, x_j^{(3)})$, $(x_j^{(3)}, x_j^{(2)})$, and $(x_j^{(2)}, x_j^{(1)})$.
	Finally, we add vertices $x_{j,j+1}^{(l)}$ and $x_{j,j+1}^{(r)}$ for every $0 \leq j \leq n$ to connect the aforementioned segments.
	This is done using edges $(x_j^{(3)}, x_{j,j+1}^{(r)})$, $(x_{j,j+1}^{(r)}, x_{j+1}^{(1)})$, $(x_{j+1}^{(1)}, x_{j,j+1}^{(l)})$, and $(x_{j,j+1}^{(l)}, x_{j}^{(3)})$.
	W.l.o.g., we can assume that all edges have length 1 and the successor functions $s_v(e) := \delta^{out}(v)$ do not restrict any movement. Moreover, we define one station 
	\begin{equation}
		S := \bigcup_{j=1}^{m} \left\{(x_j^{(1)}, x_j^{(2)}), (x_j^{(2)}, x_j^{(3)}), (x_j^{(3)}, x_j^{(2)}), (x_j^{(2)}, x_j^{(1)})\right\}.
	\end{equation}
	
	The resulting network is planar and consists of
	\begin{equation}
		2m + 2 + 5n + 2(n+1) = 2m + 7n + 4 = \mathcal{O} (n+m)
	\end{equation}
	nodes and
	\begin{equation}
		2m + 8n + 4(n+1) = 2m + 12n + 4 = \mathcal{O} (n+m)
	\end{equation}
	edges. In particular, the problem size has no exponential blow-up, and the reduction is polynomial.
	
	As an example consider the monotone 3-SAT-($\leq 2, \leq 2$) instance \mbox{$\bigwedge\limits_{i=1}^3 C_i$} with
	\begin{align}
		C_1 & = x_1 \vee x_3 \vee x_4, & C_2 & = x_2 \vee x_3 \vee x_4, & C_3 & = \neg x_1 \vee \neg x_2 \vee \neg x_4. \label{eqn:proof_example_sat}
	\end{align}
	Figure~\ref{fig:proof-example} shows the graph resulting from Equation~\ref{eqn:proof_example_sat} together with respective trains and routes, which we add in the next proof step.
	Moreover, the station $S$ is marked in orange.
	
	\begin{figure*}[!t]
		\centering
		\includegraphics[width=\textwidth]{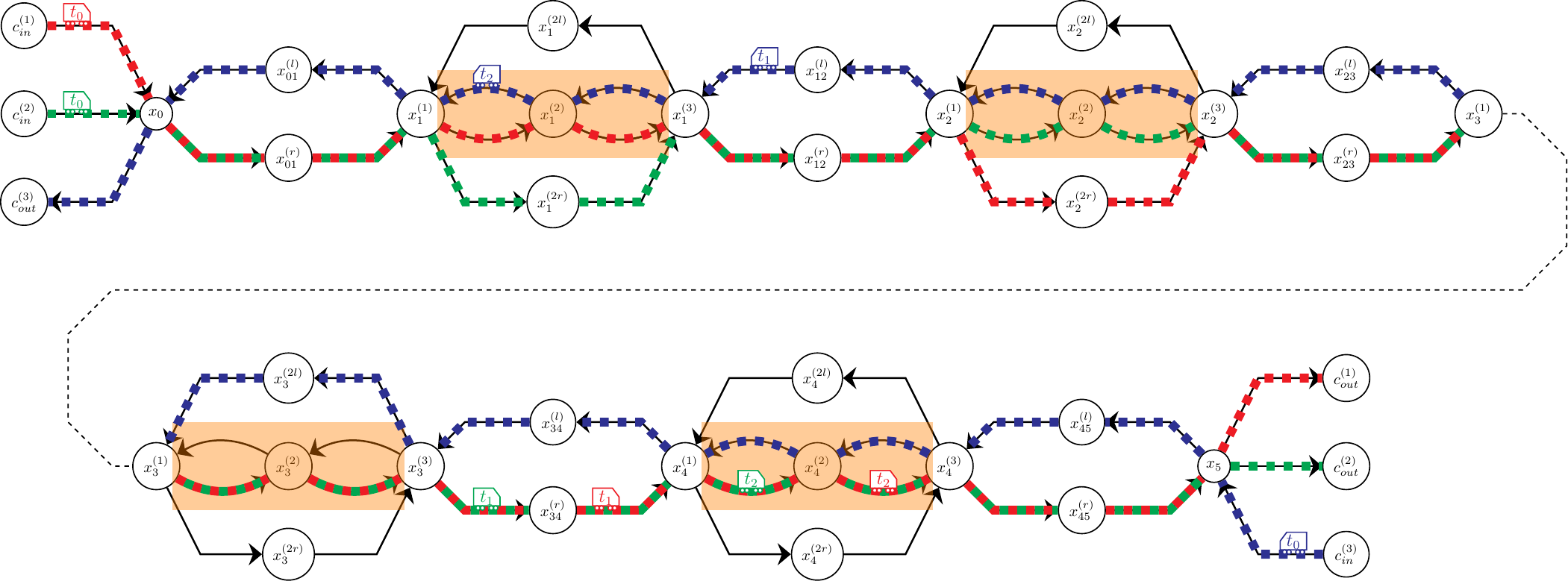}
		\caption{Reduction from Monotone 3-SAT-($\leq 2, \leq 2$) to TR-VER}
		\label{fig:proof-example}
	\end{figure*}
	Finally, we add one train $\tr_i$ (of length 1 and with maximal steed 1) per clause $C_i$ for every $1 \leq i \leq m$.
	For $1 \leq i \leq k$, the trains use the following edges
	\begin{enumerate}
		\item $c_{in}^{(i)} \to x_0 \to x_{0,1}^{(r)}$
		\item For every $j = 1,2,\ldots,n$ they traverse
		\begin{enumerate}
			\item If $x_j \in C_i$, then $x_{j-1,j}^{(r)} \to x_j^{(1)} \to x_j^{(2)} \to x_j^{(3)} \to x_{j, j+1}^{(r)}$
			\item If $x_j \not\in C_i$, then $x_{j-1,j}^{(r)} \to x_j^{(1)} \to x_j^{(2r)} \to x_j^{(3)} \to x_{j, j+1}^{(r)}$
		\end{enumerate}
		\item  Finally, $x_{n,n+1}^{(r)} \to x_{n+1} \to c_{out}^{(i)}$
	\end{enumerate}
	For $k+1 \leq i \leq m$, trains travel in the opposite direction, i.e.,
	\begin{enumerate}
		\item $c_{in}^{(i)} \to x_{n+1} \to x_{n,n+1}^{(l)}$
		\item For every $j = n,n-1,\ldots,1$ they traverse
		\begin{enumerate}
			\item If $x_j \in C_i$, then $x_{j,j+1}^{(l)} \to x_j^{(3)} \to x_j^{(2)} \to x_j^{(1)} \to x_{j-1, j}^{(l)}$
			\item If $x_j \not\in C_i$, then $x_{j,j+1}^{(l)} \to x_j^{(3)} \to x_j^{(2l)} \to x_j^{(1)} \to x_{j-1, j}^{(l)}$
		\end{enumerate}
		\item  Finally, $x_{0,1}^{(l)} \to x_{0} \to c_{out}^{(i)}$
	\end{enumerate}
	Every train's route consists of $2 + 4n + 2 = 4(n+1) =: T_n$ edges. They are marked in Figure~\ref{fig:proof-example} accordingly.
	
	All trains shall enter at time $t=0$ at their respective entry node $c_{in}^{(i)}$.
	They required to stop in $S$ from time $t = m \cdot T_n + 4$ to $t = m \cdot T_n + 6$, and shall leave the network at time $t = m \cdot T_n + 6 + m \cdot T_n = 2mT_n + 6$ at their respective exit node $c_{out}^{(i)}$.
	Since $2mT_n + 6 = \mathcal{O}(m \cdot n)$, all specified times are still polynomial in the input size.
	
	Assume that the given monotone 3-SAT-($\leq 2, \leq 2$) instance is feasible.
	From any particular solution, define $A := \{j \colon x_j \text{ is true}\}$ and $A^C := \{1, \ldots, m\} - A = \{j \colon x_j \text{ is false}\}$.
	For $1 \leq i \leq k$, define $\tilde{x}(i) := \min \{j \colon j \in A \text{ and } x_j \in C_i\}$.
	For $k+1 \leq i \leq m$, we have $\tilde{x}(i) := \min \{j \colon j \in A^C \text{ and } \neg x_j \in C_i\}$.
	The order in which trains traverse shared edges is not predefined; hence, w.l.o.g. assume that $\tilde{x}(i) \geq \tilde{x}(i+1)$ for every $1 \leq i \leq k-1$ and $\tilde{x}(i) \leq \tilde{x}(i+1)$ for every $k+1 \leq i \leq m-1$.
	Otherwise, we can rename the trains at this point.
	For $1 \leq i \leq m$, we let $\tr_i$ stop on $(c_{in}^{(1)}, x_0)$ until $t = (i-1) \cdot T_n + 1$.
	Afterwards, trains $i = 1,\ldots,k$ travel at maximal speed all the way to $(x_{\tilde{x}(i)-1, \tilde{x}(i)}^{(r)}, x_{\tilde{x}(i)}^{(1)})$ or $(x_{\tilde{x}(i)-1}^{(3)}), x_{\tilde{x}(i)-1, \tilde{x}(i)}^{(r)})$ if the previous edge is already occupied.
	Because a literal appears in only two clauses, it cannot happen that both edges are already occupied.
	Similarly, for trains traveling in the reverse direction ($i = k+1,\ldots,m$), travel at maximal speed all the way to $(x_{\tilde{x}(i), \tilde{x}(i)+1}^{(l)}, x_{\tilde{x}(i)}^{(3)})$ or $(x_{\tilde{x}(i)+1}^{(1)}), x_{\tilde{x}(i), \tilde{x}(i)+1}^{(l)})$ if the previous edge is already occupied.
	Again, because a literal appears in only two clauses, it cannot happen that both edges are already occupied.
	All trains will arrive latest at time $t = i \cdot T_n$.
	At time $t = m \cdot T_n$, they can all traverse two edges into stop $S$ within at most four time steps.
	Finally, they can proceed in the same order to the respective exit vertices in, again, at most $m \cdot T_n$ time steps and wait on their last edge without affecting other trains.
	There are no conflicts because of the chosen order and feasibility of the monotone 3-SAT-($\leq 2, \leq 2$).
	In particular, this constitutes a feasible solution to the corresponding TR-VER instance.
	In Figure~\ref{fig:proof-example}, the trains are depicted at $t_0 = 1$, $t_1 = m \cdot T_n$, and $t_2 = m \cdot T_n + 5$.
	This setting corresponds to setting $x_1$ to false and $x_4$ to true.
	
	On the other hand, if we are given a feasible timing of the corresponding TR-VER instance, we check the positions at time $t = m \cdot T_n + 5$.
	We set $x_j$ to be true if $(x_j^{(1)}, x_j^{2})$ or $(x_j^{(2)}, x_j^{3})$ is occupied and to false if $(x_j^{(3)}, x_j^{2})$ or $(x_j^{(2)}, x_j^{1})$ is occupied.
	All other variables are chosen arbitrarily.
	Note that a variable cannot be set to true and false because then trains would collide on the opposing tracks.
	Hence, this constitutes a feasible solution to the corresponding monotone 3-SAT-($\leq 2, \leq 2$) instance by design.
	
	Combining the above, we have argued that a monotone 3-SAT-($\leq 2, \leq 2$) instance is feasible if, and only if, the corresponding TR-VER instance is.
	Hence, we have polynomially reduced Monotone 3-SAT-($\leq 2, \leq 2$) to TR-VER, proving the latter's NP-completeness.
\end{proof}

Observe that no VSS is created in a verification setting. Instead, the layout is fixed. Hence, the above results essentially carry over to fixed-block systems such as classical ETCS L2.

\subsection{Generation of VSS Layouts}\label{sec:vss_layouts}
\begin{figure*}[!t]
	\centering
	\includegraphics[width=0.95\textwidth]{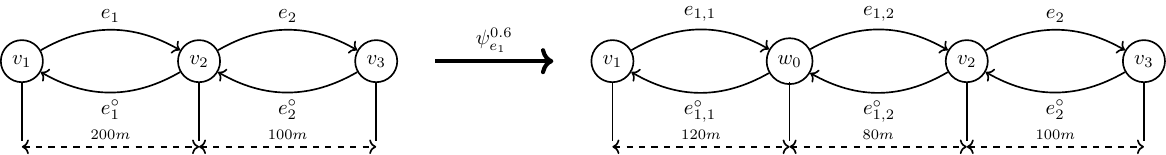}
	\caption{VSS addition operator}
	\label{fig:vss-addition-operator}
\end{figure*}

As mentioned before, the potential of hybrid train detection is that virtual subsections can be added without requiring expensive track-side detection hardware. In particular, what adding VSS sections means within our framework needs to be formalized. Namely, they correspond to inserting a VSS border vertex to separate a given edge into two or rather split both directional edges at the same position as follows.
\begin{definition}[VSS Addition Operator]
	For an edge \mbox{$e_\star=(v_1,v_2) \in E$} and a scalar \mbox{$\rho \in (0,1)$,} the \emph{VSS addition operator} $\psi_{e_\star}^{\rho}$ takes as an input a railway network \mbox{$N=(G,\len,\{s_v\}_{v \in V})$}, a border function $C_V$, and a set of stations $\mathcal{S}$. It then adds a VSS border $\rho \cdot \len(e_\star)$ away from $v_1$. More formally, it outputs a new network \mbox{$N'=(G',\len',\{s'_v\}_{v \in V'})$}, a border function $C_{V'}$, and a set of stations $\mathcal{S}'$ as follows: \\
	The vertices are expanded by a new vertex denoting a new VSS border, i.e.,
	\begin{align}
		V' & = V \cup \{v_{new}\} &
		C_{V'}(v) & = \begin{cases}
			C_V(v), & \text{ if } v \in V \\
			1, & \text{ if } v = v_{new}
		\end{cases}.
	\end{align}
	Edges $e_\star$ and possibly $e_\star^\circ$ are split at the new vertex into $e_1 := (v_1,v_{new})$ and $e_2 := (v_{new},v_2)$ plus possibly their reverse edges, i.e., using
	\begin{align}
		E^{\Delta^-} & := \{e_\star,e_\star^\circ\} \cap E &
		E^{\Delta^+} & := \begin{cases}
			\{e_1,e_2\}, & \text{ if } e_\star^\circ \not\in E \\
			\{e_1,e_2,e_1^\circ,e_2^\circ\}, & \text{ if } e_\star^\circ \in E
		\end{cases}
	\end{align}
	we have
	\begin{equation}
		E' = \left(E-E^{\Delta^-}\right) \cup E^{\Delta^+}.
	\end{equation}
	The lengths are updated according to parameter $\rho$, i.e.,
	\begin{equation}
		\len'(e) = \begin{cases}
			\len(e), & \text{ if } e \in E \\
			\rho\len(e_\star), & \text{ if } e \in \{e_1,e_1^\circ\} \\
			(1-\rho)\len(e_\star), & \text{ if } e \in \{e_2,e_2^\circ\}
		\end{cases}.
	\end{equation}
	The successor functions $\{s'_v\}_{v \in V'}$ are updated accordingly to represent the same movements in practice, i.e., using $E_1 := \{e \in \delta_E^{in}(v_1) \colon e_\star \in s_{v_1}(e)\}$ and $E_2 := \{e \in \delta_E^{in}(v_2) \colon e_\star^\circ \in s_{v_2}(e)\}$ we have
	\begin{align}
		s'_v & = s_v \text{ for all } v \in V' - \{v_1,v_2,v_{new}\} &
		s'_{v_{new}}(e) & = \begin{cases}
			\{e_2\}, & \text{ if } e = e_1 \\
			\{e_1^\circ\}, & \text{ if } e = e_2^\circ
		\end{cases}  \\
		s'_{v_1}(e) & = \begin{cases}
			(s_{v_1}(e) - \{e_\star\})\cup \{e_1\}, & \text{ if } e \in E_1 \\
			s_{v_1}(e_\star^\circ), & \text{ if } e = e_1^\circ \\
			s_{v_1}(e), & \text{ otherwise}
		\end{cases} &
		s'_{v_2}(e) & = \begin{cases}
			(s_{v_2}(e) - \{e_\star\})\cup \{e_2^\circ\}, & \text{ if } e \in E_2 \\
			s_{v_2}(e_\star), & \text{ if } e = e_2 \\
			s_{v_2}(e), & \text{ otherwise}
		\end{cases}
	\end{align}
	If a station includes $e_\star$ or its reverse, it is updated accordingly, i.e., using \mbox{$f\left(e_\star\right) := \{e_1,e_2\}$}, \mbox{$f\left(e_\star^\circ\right) := \{e_1^\circ,e_2^\circ\}$}, and
	\begin{align}
		\Delta^-(S) & := \left\{e_\star,e_\star^\circ\right\} \cap S &
		\Delta^+(S) & := \bigcup_{e \in \Delta^-(S)} f(e)
	\end{align}
	we have
	\begin{equation}
		\mathcal{S}' = \left\{\left(S-\Delta^-(S)\right) \cup \Delta^+(S) \colon S \in \mathcal{S}\right\}.
	\end{equation}
	We denote by $\Psi$ the set of all VSS addition operators.
\end{definition}
\begin{example}
	Consider the graph depicted in Figure~\ref{fig:vss-addition-operator}.
	Assume we want to separate the track between $v_1$ and $v_2$ into two VSS sections corresponding to adding a VSS border, say $w_0$.
	The first section has a length of $0.6 \cdot 200m = 120m$, the latter of $80m$.
	The transformation is shown in Figure~\ref{fig:vss-addition-operator} and corresponds to the operator $\psi_{e_1}^{0.6}$, which is equivalent to $\psi_{e_1^\circ}^{0.4}$.
\end{example}
Some elements (e.g., turnouts) require special treatment and classical train separation.
In our formulation, this can be achieved by marking some of the TTD sections as \emph{unbreakable}.
It is then forbidden to apply the VSS addition operator on an edge within any unbreakable TTD section.
For example, any TTD section containing a turnout could be marked in such a way.
By doing this, the control logic around switches would not change, while at the same time allowing optimization using VSS on all permitted tracks (only).

\begin{figure*}[!t]
	\centering
	\includegraphics[width=0.8\textwidth]{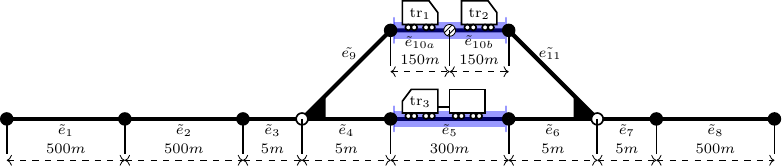}
	\caption{Generation of VSS layout $(t=300)$}
	\label{fig:vss-generation}
\end{figure*}

By leaving the TTDs fixed and only adding VSS sections, we can change the block structure at a low cost. Because of this, it might be possible to change the virtual subsections when a new train schedule should be implemented. Hence, we can somewhat \enquote{overfit} to the current timetable requests without relying solely on schedule-independent KPIs (e.g., headway). This leads to the following design task of generating VSS layouts.
\begin{problem}
	\problemtitle{Generation of VSS Layout (VSS-GEN)}
	\probleminput{A railway network $N=(G,\len,\{s_v\}_{v \in V})$; border function $C_V \colon V \to \{0,1,2\}$; trains $\tr_1,\ldots,\tr_m$; a list of timetable requests for $\tr_1$ to $\tr_m$.}
	\problemquestion{Determine a minimal number of VSS addition operators $\psi^{(1)},\ldots,\psi^{(k)} \in \Psi$ (and train routes $R^{(\tr_1)},\ldots,R^{(\tr_m)}$) such that on \mbox{$\psi^{(k)}\circ \ldots \circ \psi^{(1)}(\cdot)$} the VSS condition is fulfilled and all trains respect their speed and timetable requests or output that no such operators exist.}
\end{problem}
If possible, we try to keep the number of additional virtual subsections small to keep it tractable for the control system during operation. In particular, we do not require a hard optimum because one section more or less will likely not pose any additional costs. On the other hand, it might be interesting to still have a minimal number of subsections to make given timetable requests realizable. In a later stage, one could then add more subsections to improve robustness; if the generated VSS layout from this design task is minimal, there might be more space to add subsections to improve robustness.
\begin{example}
	Again, consider the setting from Example~\ref{ex:verification} and Figure~\ref{fig:infeasible-verification}. If we separate $\tilde{e}_{10}$ into two virtual subsections by including a VSS border (\begin{tikzpicture} \draw[preaction={fill, white}, pattern=north east lines, pattern color=black] (0,0) circle (0.1); \end{tikzpicture}), the train station $S_0$ has three blocks.
	In fact, we can now route the trains as shown in Figure~\ref{fig:vss-generation}, i.e., in the end, trains $\tr_1$ and $\tr_2$ can occupy the newly created subsections and train $\tr_3$ the other track. Hence, by adding one VSS border, we have made the previously infeasible timetable requests realizable. Note that we could also have added additional subsections, e.g., split track $\tilde{e}_5$ and have $\tr_3$ occupy two blocks in the end. However, as said, our objective is to add rather few VSS borders, and one additional VSS border suffices to make the timetable requests realizable.
\end{example}
Since VSS-GEN is not a decision problem, NP-completeness is not defined in this context. However, NP-hardness can be shown by relating it to one of the previously defined verification tasks, where we use the more general definition that a problem is NP-hard if it can be reduced from some NP-complete problem, hence, not requiring the NP-hard problem to be of decision type \cite{Leeuwen1990}.
\begin{theorem}
	VSS-GEN is NP-hard even if all routes and exact stopping times in train stations are fixed and the railway network is planar.
\end{theorem}
\begin{proof}
	Consider an instance of \mbox{TR-VER}. Assume we are given an oracle to solve any instance of \mbox{VSS-GEN}. The given \mbox{TR-VER} instance constitutes valid input to \mbox{VSS-GEN}. Let $k^\star$ be the number of VSS addition operators used in an optimal solution of \mbox{VSS-GEN} on the respective input or let \mbox{$k^\star = \infty$} if no feasible solution exists. Then, the solution to the given \mbox{TR-VER} instance is \enquote{yes} if, and only if, \mbox{$k^\star = 0$}. In particular, we have polynomially reduced \mbox{TR-VER} to \mbox{VSS-GEN}. The claim follows directly from Theorem~\ref{thm:complexity-verification}.
\end{proof}
\begin{remark}
	Instead of adding additional vertices in a continuous setting using VSS addition operators, we might also consider a discretized version, which might be easier to handle computationally. In this setting, we would require the railway network $N$ to consist of all nodes that could possibly become a VSS border. Adding a VSS border is then equivalent to changing $C_V(v)$ from $0$ to $1$ for some $v \in V$ with $|\Gamma(v)| \leq 2$, where $\Gamma$ denotes the neighborhood (ignoring the direction of edges).
\end{remark}

\subsection{Schedule Optimization Using the Potential of VSS}\label{sec:schedule}
In practice, minimizing VSS sections might not be of great interest.
Instead, one might want to optimize the operational outcome, e.g., by minimizing travel times.
Intuitively, one might think that this is an (computationally) easy problem, because the \enquote{best} layout is creating very small VSS (e.g., every meter or even less).
In that case, trains would basically operate as if a moving block control system were implemented.
However, according to industry experts, such an implementation would not be feasible because every virtual block has to be specified within the interlocking and the \emph{Radio Block Center}~(RBC).
These components must fulfill the \emph{safety integrity level}~(SIL)~4 and are time-consuming and costly to develop and maintain.
Because many such systems were developed and approved for classical control systems, the address space for block sections has limited capacity.

Thus, in practice, there is a significant upper limit on how many VSS can be defined, which should be considered within the planning process.
Hence, the question arises on strategically placing these VSS to optimize the operational outcome.
In particular, a proposed solution of placing a VSS every meter is likely not feasible.

\begin{problem}
	\problemtitle{Schedule Optimization (VSS-OPT)}
	\probleminput{A railway network $N=(G,\len,\{s_v\}_{v \in V})$; border function $C_V \colon V \to \{0,1,2\}$; trains $\tr_1,\ldots,\tr_m$; a list of timetable requests for $\tr_1$ to $\tr_m$; an upper bound $K_{max}$.}
	\problemquestion{Determine VSS addition operators $\psi^{(1)},\ldots,\psi^{(k)} \in \Psi$ (and train routes $R^{(\tr_1)},\ldots,R^{(\tr_m)}$) such that $k \leq K_{max}$, on \mbox{$\psi^{(k)}\circ \ldots \circ \psi^{(1)}(\cdot)$} the VSS condition is fulfilled and all trains respect their speed and timetable requests or output that no such operators exist.
		Among all feasible solutions, the one with minimal travel times is returned. As an objective, one might, e.g., minimize the (weighted) sum of travel times or the longest travel time.}
\end{problem}

Another operational objective might be to maximize capacity.
Assume, e.g., one plans a mixed railway network on which passenger trains should operate, adhering to certain scheduling constraints.
On the other hand, there are additional requests for freight trains running through the network.
In this setting, one might want to maximize the number of freight trains without negatively affecting the timetable requests of passenger trains by using the full potential of VSS sections.

\begin{problem}
	\problemtitle{Capacity Optimization (VSS-CAP)}
	\probleminput{A railway network $N=(G,\len,\{s_v\}_{v \in V})$; border function $C_V \colon V \to \{0,1,2\}$; trains $\tr_1,\ldots,\tr_{m_1}$ and $\tilde{\tr}_1,\ldots,\tilde{\tr}_{m_2}$; a list of timetable requests for every train; an upper bound $K_{max}$.}
	\problemquestion{Determine VSS addition operators $\psi^{(1)},\ldots,\psi^{(k)} \in \Psi$, a subset $\mathcal{R} \subseteq \{1,\ldots,m_2\}$, train routes $R^{(\tr_1)},\ldots,R^{(\tr_m)}$ and $R^{(\tilde{\tr}_j)}$ ($j \in \mathcal{R}$) such that $k \leq K_{max}$, on \mbox{$\psi^{(k)}\circ \ldots \circ \psi^{(1)}(\cdot)$} the VSS condition is fulfilled and all routed trains respect their speed and timetable requests or output that no such operators exist.
		Among all feasible solutions, the one maximizing $|\mathcal{R}|$ (i.e., the number of routed optional trains).}
\end{problem}
\begin{theorem}
	VSS-OPT and VSS-CAP are NP-hard even if all (potential) routes and exact stopping times in train stations are fixed, and the railway network is planar.
\end{theorem}
\begin{proof}
	Consider an instance of \mbox{TR-VER}.
	Assume we are given an oracle to solve any instance of \mbox{VSS-OPT}.
	The given \mbox{TR-VER} instance constitutes valid input to \mbox{VSS-OPT}.
	Let $t^\star$ be the optimal travel time objective or \mbox{$t^\star = \infty$} if no feasible solution exists.
	Observe that it is easy to deduce a theoretical upper bound on $t^\star$ in the feasible case, say $\overline{t}$, from the latest possible exit times $\tu_{out}^{tr_1}, \ldots , \tu_{out}^{tr_m}$, e.g., $\overline{t} = \tu_{out}^{tr_1} + \ldots + \tu_{out}^{tr_m}$ if the objective is given by the sum of travel times.
	Then, the solution to the given \mbox{TR-VER} instance is \enquote{yes} if, and only if, \mbox{$t^\star \neq \infty$}, i.e., $t^\star \leq \overline{t}$.
	
	The given \mbox{TR-VER} instance also constitutes valid input to \mbox{VSS-CAP}.
	Let $r^\star = |\mathcal{R^\star}|$ be the maximal number of optional trains $\tilde{\tr}_1, \ldots, \tilde{\tr}_{m_2}$ that are routed or $r^\star = - \infty$ if the problem is infeasible even for $\mathcal{R} = \emptyset$.
	Then, the solution to the given \mbox{TR-VER} instance is \enquote{yes} if, and only if, \mbox{$r^\star \neq \infty$}, i.e., $r^\star \leq m_2$.
	
	In particular, we have polynomially reduced \mbox{TR-VER} to \mbox{VSS-OPT} and \mbox{VSS-CAP}. The claim follows directly from Theorem~\ref{thm:complexity-verification}.
\end{proof}

\section{Conclusions}\label{sec:concl}
In this paper, we formalized important properties of ETCS~L2~HTD and corresponding design tasks. Moreover, we argued that the framework could eventually also model more realistic train movements as additional constraints by stating some examples of possible extensions in Section~\ref{sec:preliminaries}. While the verification tasks in Section~\ref{sec:ver} are analog to older level systems (since the block layout is fixed), the two latter tasks (Sections~\ref{sec:vss_layouts}~and~\ref{sec:schedule}) make use of the new freedom provided by virtual subsections. Furthermore, we proved that these tasks are NP-hard (or NP-complete, respectively). By this, we provided a theoretical foundation for further developing corresponding design methods for ETCS~L2~HTD.
These methods will be integrated into the Munich Train Control Toolkit (MTCT) hosted on GitHub (\url{https://github.com/cda-tum/mtct}).

The results confirm that previously proposed heuristics and exact methods (presented, e.g., in~\cite{Wille2021, Peham2022}), which only generate non-optimal results or have a severely limited scalability, respectively, do not suffer from insufficient methodology but complexity reasons. 
Overall, for the first time, a solid basis for developing (ideally scalable) design automation solutions incorporating the freedom allowed by virtual subsections within ETCS~L2~HTD is provided. These methods might be a valuable addition to the already existing work on fixed-block control systems summarized in Section~\ref{sec:intro}. It will also be interesting to see how these methods tailored to hybrid train detection systems perform compared to previously designed methods that do not consider the specific timetable requests for generating signaling layouts.
	
	\bibliographystyle{elsarticle-num}
	{%
		\bibliography{lit-etcs}%
	}

\end{document}